\documentclass[journal]{IEEEtran}
\usepackage{amsmath,amsthm,amssymb,bm,tikz,cite,cases}
\usepackage{multirow}
\usepackage{bbm}
\usepackage{algpseudocode}
\usepackage{algorithm}
\usepackage[hidelinks]{hyperref}

\newcommand{\Aut}[1]{\mathbb{A}_{#1}}
\DeclareMathOperator{\RM}{RM}
\DeclareMathOperator{\eval}{eval}
\DeclareMathOperator{\hard}{b}
\DeclareMathOperator{\GMCdec}{\Gamma}
\newcommand{\GMC}[2]{\GMCdec_{#1, #2}}
\DeclareMathOperator{\permGMCdec}{L}
\newcommand{\permGMC}[3]{\permGMCdec_{#1, #2}^{#3}}
\DeclareMathOperator{\SCL}{\Xi}
\DeclareMathOperator{\CA}{\Psi}
\DeclareMathOperator{\XR}{{\overline{\mathbb{R}}}}
\newcommand{\Tree}[2]{\mathbb{T}_{#1,#2}}
\newcommand{\cost}{\mathcal{L}}
\DeclareMathOperator{\lsigmoid}{\eta}
\DeclareMathOperator{\id}{\mathsf{id}}

\DeclareRobustCommand{\rchi}{{\mathpalette\irchi\relax}}
\newcommand{\irchi}[2]{\raisebox{\depth}{$#1\chi$}}
\DeclareMathOperator{\Complexity}{\rchi}
\newcommand{\COMP}[1]{\Complexity_{\text{#1}}}
\newcommand{\lin}{\ell_{\text{in}}}
\newcommand{\lout}{\ell_{\text{out}}}
\newcommand{\lmax}{\ell_{\max}}

\begin{document}

\title{Constituent Automorphism Decoding of Reed--Muller Codes}
\author{Yicheng Qu, Amir Tasbihi, and Frank R.~Kschischang\thanks{The
authors are with the Edward S.\ Rogers Sr.\ Dept.\ of Electrical and
Computer Engineering, University of Toronto, 10 King's College Road,
Toronto, Ontario M5S 3G4, Canada.  Email: \texttt{eason.qu@mail.utoronto.ca},
\texttt{amir.tasbihi@mail.utoronto.ca}, \texttt{frank@ece.utoronto.ca}.  Submitted for
publication on July 19th, 2024.}}

\maketitle

\begin{abstract}
Automorphism-ensemble decoding is applied to the Plotkin constituents of
Reed--Muller codes, resulting in a new soft-decision decoding algorithm
with state-of-the-art performance versus complexity trade-offs.
\end{abstract}


\section{Introduction}
\label{sec::intro}

\IEEEPARstart{T}{his} paper introduces the \emph{constituent
automorphism} (CA) decoder, a new soft-decision decoding algorithm for
Reed--Muller (RM) codes.  The new algorithm, a variant of automorphism
ensemble (AE) decoding \cite{AE-SC}, exploits the recursive Plotkin
structure of RM codes, applying AE decoding to constituent RM codes
located at various levels in the resulting decoding tree.  By carefully
adjusting the number of automorphisms applied to particular constituent
decoders, CA decoding results in better performance versus complexity
trade-offs than other state-of-the-art decoding algorithms such as
successive cancellation list (SCL) decoding~\cite{recurse-list} and AE
decoding itself \cite{AE-SC} (where automorphisms are applied only at
the root of the decoding tree).

Although Reed--Muller codes are among the earliest families of codes
introduced in coding theory \cite{Muller,Reed}, contemporary interest in
them arises from (a) their excellent soft-decision decoding performance
at short block lengths, (b) their close connection to polar codes, (c)
the relatively recent discovery that RM codes can be capacity-achieving
in certain situations, and (d) the fact that RM codes arise in
connection with various problems in theoretical computer science; see
\cite{abbe2023reed} for an excellent review.  In~\cite{GMC}, Schnabl and
Bossert view RM codes as generalized multiple concatenated (GMC) codes,
introducing the first soft-decision RM decoding algorithm---herein
referred to as the GMC decoder---by exploiting their recursive Plotkin
structure to successively decode their constituent codes.  In the
literature on polar codes~\cite{channel-polarization}, the GMC decoder
is also often referred to as a \emph{successive cancellation} (SC)
decoder.  In \cite{recurse-list}, Dumer and Shabunov show that better RM
soft-decision decoding performance can be obtained for the binary-input
additive white Gaussian noise (BI-AWGN) channel by combining GMC
decoding with list-decoding.  The resulting successive-cancellation list
(SCL) decoder was extended to polar codes in \cite{SCL}.

The idea of exploiting code automorphisms for decoding dates
back to the 1960s \cite{Prange_TheUseOf_1962,permutationf-of-systematic-codes}.
Recent interest in
automorphism-based decoders for RM codes~\cite{kamenev2019anew,
ivanov2019permutation, AE-SC} stems from
their ability to approach
maximum-likelihood (ML) decoding performance with a reasonable
complexity.
The general idea is to permute the
received vector by the elements of a set $\mathcal{E}$ of code
automorphisms, then to decode the permuted versions in parallel via a
simple sub-optimal \emph{elementary decoder}. The inverse automorphism
is then applied to each decoded word to arrive at a list of
$|\mathcal{E}|$ (not necessarily distinct) decoding candidates, where
$|\mathcal{E}|$ denotes the number of elements in $\mathcal{E}$.  The
decoder then selects the most likely codeword from the candidate list.
While the automorphism-based decoder is approximately $|\mathcal{E}|$
times computationally more complex than its elementary decoder, its
inherent parallelism makes it very hardware-friendly \cite{AE-with-PM},
which may be seen as an advantage of AE decoding compared with SCL
decoding.  The key idea of this paper is to apply AE decoding at the
level of the Plotkin constituent codes of a Reed--Muller code.  This
maintains a high degree of parallelism in the decoding algorithm and, as
we demonstrate, it can result in performance-complexity benefits.

The remainder of this paper is organized as follows.  Various
coding-theoretic preliminaries needed to understand the rest of the
paper are reviewed in Sec.~\ref{sec::background}.  The CA decoder is
introduced in Sec.~\ref{sec::CA}, and its complexity is analyzed in
Sec.~\ref{sec::complexity}.   Simulation results showing performance and
complexity tradeoffs are given in Sec.~\ref{sec::results}, along with
comparisons to other state-of-the-art RM decoders.  Some brief
concluding remarks are given in Sec.~\ref{sec::conclusions}.

Throughout this paper, the real numbers are denoted by $\mathbb{R}$, the
integers are denoted by $\mathbb{Z}$, and the two-element finite field
$\{ 0, 1 \}$ is denoted by $\mathbb{F}_2$.  The extended real numbers
$\mathbb{R} \cup \{ -\infty, +\infty \}$ are denoted as
$\XR$.  The (modulo-two) addition operation in
$\mathbb{F}_2$, equivalent to an exclusive-OR (XOR) operation, is
denoted as $\oplus$. The \emph{soft XOR} of two extended real numbers
$a, b \in \XR$, denoted as $a \boxplus b$, is given as
\[
a \boxplus b \triangleq 2\tanh^{-1}\left( \tanh \left(
\frac{a}{2}\right) \tanh \left( \frac{b}{2}\right) \right),
\]
where $\tanh( \pm \infty) = \pm 1$, and where the symbol `$\triangleq$'
signifies that the right-hand side is the definition of the left-hand
side.  The \emph{negative log-sigmoid} of
$x \in \XR$ is given as
\begin{equation}
\lsigmoid(x) \triangleq \ln(1 + \exp(-x)),
\label{eq:lsigmoid}
\end{equation}
where $\lsigmoid(-\infty) = \infty$ and $\lsigmoid(\infty) = 0$.
For any integer $k$,
\[
\mathbb{Z}^{\geq k}\triangleq \{ k, k+1, \ldots \} \text{ and }
\mathbb{Z}^{\leq k}\triangleq \{ \ldots, k-1, k \}.
\]
For any $n \in \mathbb{Z}^{\geq 0}$, $[n] \triangleq \{ 0, 1, \ldots,
n-1 \}$, where $[0]$ is the empty set.  The mapping $(-1)^{(\cdot)}:
\mathbb{F}_2 \to \mathbb{R}$ is defined, for $b \in \mathbb{F}_2$, as
\[
(-1)^b \triangleq \begin{cases} 1, & \text{if }b=0; \\
                      -1, & \text{if }b=1. \end{cases}
\]
For any predicate $p$, $\mathbbm{1}_p$ denotes the indicator function 
\[
\mathbbm{1}_p \triangleq \begin{cases}
1, & \text{if } p \text{ is true};\\
0, & \text{otherwise}.
\end{cases}
\]
The codomain of $\mathbbm{1}_p$ may be $\XR$ or
$\mathbb{F}_2$ as context dictates.  Vectors (over
$\XR$ or $\mathbb{F}_2$) are denoted with boldface
lower-case letters.  A vector of length $n$, \emph{i.e.}, a vector
having $n$ components, is referred to as an $n$-vector.  The $i$th
component of an $n$-vector $\mathbf{v}$ is denoted $\mathbf{v}[i]$,
where $i \in [n]$.  The empty vector $\varnothing \triangleq ()$ has
zero length and no components.  The \emph{concatenation} of an
$n_1$-vector $\mathbf{u}$ and an $n_2$-vector $\mathbf{v}$, in that
order, is the $(n_1+n_2)$-vector
\[
( \mathbf{u} \mid \mathbf{v} ) \triangleq
( \mathbf{u}[0], \ldots, \mathbf{u}[n_1-1],
  \mathbf{v}[0], \ldots, \mathbf{v}[n_2-1] ).
\]
Generally $(\mathbf{u} \mid \mathbf{v} ) \neq (\mathbf{v} \mid
\mathbf{u})$, although $( \mathbf{u} \mid \varnothing) = (\varnothing
\mid \mathbf{u}) = \mathbf{u}$.  The all-zero and all-one $n$-vectors
over $\mathbb{F}_2$ are denoted as $\bm{0}_n$ and $\bm{1}_n$,
respectively, where the subscript may be omitted if the length is clear
from the context.  For any binary vector $\mathbf{b}$ and any real vector
$\bm{\lambda}$ of the same length, $(-1)^{\mathbf{b}} \bm{\lambda}$
denotes a real vector whose $i$th component is $(-1)^{\mathbf{b}[i]}
\bm{\lambda}[i]$.  For any $n \in \mathbb{Z}^{\geq 0}$, $\left((a_i,
b_i)_{i \in [n]}\right)$ denotes the $n$-tuple of $2$-tuples $((a_0,
b_0), (a_1, b_1), \ldots, (a_{n-1}, b_{n-1}))$, where the $a_i$'s and
$b_i$'s are arbitrary mathematical objects. This notation may be generalized to
$n$-tuples of $m$-tuples in the obvious way.  For any $m \in
\mathbb{Z}^{\geq 0}$, the ring of $m$-variate binary polynomials with
indeterminates $x_0, \ldots, x_{m-1}$ is denoted by $\mathbb{F}_2[x_0,
\ldots, x_{m-1}]$.  When $m=0$ we have $\mathbb{F}_2[\,]
\triangleq \mathbb{F}_2$.  For any $r \in \mathbb{Z}$,
$\mathbb{F}_2^{\leq r} [x_0, \ldots, x_{m-1}]$ denotes the set of
polynomials in $\mathbb{F}_2[x_0, \ldots, x_{m-1}]$ whose total degree
is at most $r$.  The total degree of the zero polynomial $0$ is taken as
$-\infty$; thus for any $r < 0$, $\mathbb{F}_2^{\leq
r}[x_0,\ldots,x_{m-1}] = \{ 0 \}$.

\section{Preliminaries}\label{sec::background}

In this section we briefly review various coding-theoretic
concepts needed to understand the remainder of the paper.

\subsection{Channels and Decoders}

Throughout this paper, we assume the use of a binary-input memoryless
channel with input alphabet $\mathbb{F}_2$, output alphabet
$\mathcal{Y}$, and channel law given, for $x \in \mathbb{F}_2$ and $y
\in \mathcal{Y}$, as $W(y \mid x)$, where $W$ is either a conditional
probability mass function or conditional probability density function
according to whether $\mathcal{Y}$ is discrete or continuous.  Without
loss of generality, we assume that $\mathcal{Y}$ is chosen so that $W(y
\mid 0) \neq  0$ or $W(y \mid 1) \neq 0$ for all $y \in \mathcal{Y}$.

A standard example of such a channel is the binary symmetric channel
with crossover probability $p$, where $\mathcal{Y} = \mathbb{F}_2$
and $W(1\mid 0)=W(0\mid 1)=p$.  Another example is the binary erasure
channel (BEC) with erasure probability $\epsilon$, where $\mathcal{Y} =
\mathbb{F}_2 \cup \{ e \}$ and $W(e \mid 0) = W(e \mid 1) = \epsilon$
and $W(0 \mid 1)=W(1 \mid 0) = 0$.  Another standard example, and the
one we use to present simulation results, is the BI-AWGN with noise
variance $\sigma^2$, where $\mathcal{Y} = \mathbb{R}$, with $W(y \mid 0)
= \mathcal{N}(1,\sigma^2)$ and $W(y \mid 1) = \mathcal{N}(-1,\sigma^2)$,
where $\mathcal{N}(m,\sigma^2)$ denotes a Gaussian density function with
parameters $m$ and $\sigma^2$.  The signal-to-noise ratio (SNR)
of such a channel is $1/\sigma^2$.

Associated with each received channel output $y \in \mathcal{Y}$ is the log-likelihood
ratio (LLR) $\Lambda(y) \triangleq \ln \left( \frac{ W(y \mid 0) }{ W(y
\mid 1)} \right) \in \XR$, where $\Lambda(y) =
-\infty$ if $W(y \mid 0) = 0$ and $\Lambda(y) = +\infty$ if $W(y \mid 1)
= 0$.  Infinite LLRs occur, for example, for unerased symbols received
at the output of the BEC.  The \emph{hard decision} associated with an
LLR value $\lambda = \Lambda(y)$, for some channel output $y$, is given
by the function $\hard: \XR \to \mathbb{F}_2$, where
$\hard(\lambda) \triangleq \mathbbm{1}_{\lambda < 0}$.

Suppose that a binary codeword of length $n$ is transmitted.  Based on
the corresponding channel output $\mathbf{y} \in \mathcal{Y}^n$, we
assume that the receiver produces the log-likelihood-ratio (LLR) vector
$\bm{\lambda} \in \XR^n$, where $\bm{\lambda}[i]  =
\Lambda(\mathbf{y}[i])$ for $i \in [n]$.  The LLR vector serves as the
interface between the channel and the decoder.  We also define the
vector of hard-decisions $\hard(\bm{\lambda}) \in \mathbb{F}_2^n$ as a
binary $n$-vector whose entries are hard decisions associated with
$\bm{\lambda}$ entries, \emph{i.e.}, $\hard(\bm{\lambda})[i] \triangleq
\hard(\bm{\lambda}[i])$ for $i \in [n]$.

A \emph{decoding rule} or simply \emph{decoder} for a binary code $C$ of
length $n$ is a function $D: \XR^n \to C \cup \{ F \}$
that maps an LLR vector $\bm{\lambda}$ either to a codeword of $C$ or to
the symbol $F$ (which indicates a decoding failure).  Two decoders $D_1$
and $D_2$ are \emph{the same} for a given channel if they almost surely
produce the same decoding decision, \emph{i.e.}, if
$\Pr(D_1(\bm{\lambda}) \neq D_2(\bm{\lambda})) = 0$, where
$\bm{\lambda}$ is the LLR vector corresponding to the channel output.
We write $D_1 \neq D_2$ when $D_1$ and $D_2$ are not the same.

The \emph{analog weight} $w(\mathbf{x}, \bm{\lambda})$ of a channel
input $\mathbf{x}$ with respect to an LLR vector $\bm{\lambda}$ is given
as the sum of the absolute values of those components of $\bm{\lambda}$
whose hard-decisions disagree with the corresponding component of
$\mathbf{x}$, \emph{i.e.},
\begin{equation}
w(\mathbf{x},\bm{\lambda}) \triangleq
\sum_{i:\mathbf{x}[i] \neq \hard(\bm{\lambda}[i])}
|\bm{\lambda}[i]|.
\label{eq::analog_weight}
\end{equation}
The analog weight $w(\mathbf{x},\bm{\lambda})$ measures the cost of
disagreements between $\mathbf{x}$ and $\hard(\bm{\lambda})$.  Among all
possible channel inputs, the hard-decision vector $\hard(\bm{\lambda})$ itself
has minimum possible analog weight, namely $0$;  however, $\hard(\bm{\lambda})$
may not be a codeword.  Given an LLR vector $\bm{\lambda}$,
a \emph{maximum-likelihood (ML) decoder}
for a code $C$ produces a codeword $\mathbf{v} \in C$ having least
analog weight with respect to $\bm{\lambda}$.
If codewords are equally likely to be transmitted, an ML
decoder minimizes block error rate (BLER), \emph{i.e.}, the probability
that the decoding decision disagrees with the transmitted codeword.

\subsection{Code Automorphisms}

For any positive integer $n$, a \emph{permutation} of order $n$ is a
bijection $\pi : [n] \to [n]$.  The set of all permutations of order $n$
forms a group (the \emph{symmetric group} $S_n$) under function
composition.  The group identity is denoted as $\id$.  The symmetric
group $S_n$ acts on $n$-vectors by permutation of coordinates,
\emph{i.e.}, for any $n$-vector $\mathbf{v}$ and any $\pi \in S_n$,
\[
\pi \mathbf{v} \triangleq ( \mathbf{v}[\pi(0)], \ldots ,
\mathbf{v}[\pi(n-1)] ).
\]

A permutation $\pi \in S_n$ is called an \emph{automorphism} of a code
$C$ of length $n$ if $\mathbf{c} \in C$ implies $\pi\mathbf{c} \in C$.
In other words, an automorphism of a code is a permutation which maps
codewords to codewords.  The set of all automorphisms of $C$ form a
subgroup of $S_n$ called the \emph{automorphism group} of $C$.

\subsection{Reed--Muller Codes}

Fix $m \in \mathbb{Z}^{\geq 0}$.  For any
polynomial $p = p(x_0,\ldots,x_{m-1}) \in \mathbb{F}_2[x_0, \ldots,
x_{m-1}]$, and any binary $m$-vector $\mathbf{v}$, called a
\emph{point}, let $p(\mathbf{v}) = p(\mathbf{v}[0], \ldots,
\mathbf{v}[m-1]) \in \mathbb{F}_2$ denote the \emph{evaluation} of the
polynomial $p$ at the point $\mathbf{v}$, \emph{i.e.}, the binary value
obtained when $x_i$ is substituted with $\mathbf{v}[i]$, for each $i \in
[m]$.  When $m=0$, we have $p(\varnothing) = 0$ if $p = 0$ and
$p(\varnothing)=1$ if $p=1$.  Define the \emph{evaluation map} $\eval:
\mathbb{F}_2 [x_0, \ldots, x_{m-1}] \to \mathbb{F}_2^{2^m}$ as the
function which maps the polynomial $p \in \mathbb{F} [x_0, \ldots,
x_{m-1}]$ to its evaluation at \emph{all} points of $\mathbb{F}_2^m$ in
lexicographic order, \emph{i.e.}, 
\begin{multline}
\eval(p) \triangleq \Big(
p(0,\ldots,0,0), p(0,\ldots,0,1), p(0,\ldots,1,0),\\
     \ldots, p(1,\ldots,1,1)\Big).
\label{eq:evalmap}
\end{multline}
For any $r \in \mathbb{Z}$ and any $m \in \mathbb{Z}^{\geq 0}$, the
binary Reed--Muller code of \emph{log-length} $m$ and \emph{order} $r$,
denoted as $\RM(r,m)$, is the image of $\mathbb{F}_2^{\leq
r}[x_0,\ldots,x_{m-1}]$ under the evaluation map, \emph{i.e.},
\begin{equation}
\RM(r,m) \triangleq \left\{
\eval(p): p \in 
\mathbb{F}_2^{\leq r}[x_0, \ldots, x_{m-1}]
\right\}.
\end{equation}
It can be shown (see, \emph{e.g.}, \cite[Fig.~13.4]{Macwilliams-book}), for any
$m \in \mathbb{Z}^{\geq 0}$ and any integer $r$ satisfying $0 \leq r
\leq m$, that $\RM(r,m)$ is a code of block length $n = 2^m$, dimension
\[
k(r,m) \triangleq \sum_{i=0}^r \binom{m}{i}
\]
and minimum Hamming distance $d = 2^{m-r}$.  When $r < 0$, $\RM(r,m) =
\{ \mathbf{0}_{2^m} \}$ contains only the zero codeword.  When $r=0$, $\RM(r,m)$ is the
binary repetition code of length $2^m$.  When
$r=m-1$, $\RM(m-1,m)$ is the $(2^m,2^m-1)$ single-parity check (SPC) code.
When $r \geq m$, $\RM(r,m) =
\mathbb{F}_2^{2^m}$ (the whole space of $2^m$-vectors).  These are the
\emph{trivial} RM codes.  When $0 < r < m-1$, we call $\RM(r,m)$
\emph{nontrivial}.

RM codes are one of the few algebraic code families with a
well-characterized automorphism group.  We denote the automorphism
group of $\RM(r,m)$ by $\Aut{r,m}$.  When $\RM(r,m)$ is nontrivial, it
is known \cite{abbe2023reed}\cite[Sec.~13.9]{Macwilliams-book} that
$\Aut{r,m}$ is isomorphic to the general affine group $\text{GA}(m,
\mathbb{F}_2)$.  When $\RM(r,m)$ is trivial, $\Aut{r,m}$ is equal to
$S_{2^m}$ (\emph{i.e.}, \emph{every} permutation of coordinates is an
automorphism).

Under the lexicographic order by which the evaluation map in
(\ref{eq:evalmap}) is defined, one may show that RM codes admit a
Plotkin structure~\cite{Plotkin}.  This means that any codeword
$\mathbf{c} \in \RM(r,m)$, where $m \geq 1$, can be written as 
\[
 \mathbf{c} = (\mathbf{u} \mid \mathbf{u}\oplus\mathbf{v}),
\]
where $\mathbf{u} \in \RM(r,m-1)$ and $\mathbf{v} \in \RM(r-1,m-1)$.  We
refer to $\RM(r,m-1)$ and $\RM(r-1,m-1)$ as the \emph{Plotkin
constituents} of $\RM(r,m)$.  Of course $\RM(r,m-1)$ and $\RM(r-1,m-1)$
are themselves RM codes, having their own Plotkin constituents.

\begin{figure}[t]
\centering
\includegraphics[scale=0.6666]{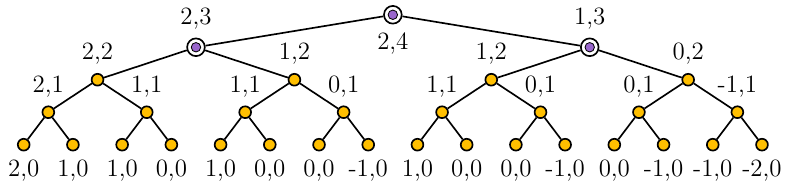}
\caption{Illustrated is $\Tree{2}{4}$. The label of each
vertex is $(r,m)$, for some $r\in \mathbb{Z}$ and $m \in \mathbb{Z}^{\geq
0}$, where for brevity the parentheses are omitted.  The circled
vertices form $\Tree{2}{4}^{\mathcal{A}^{\ast}}$, the GMC decoding tree
for $\RM(2,4)$ with atom set $\mathcal{A}^{\ast}$ in (\ref{eq:atomset}).}
\label{fig:tree}
\end{figure}

The relationships among the various Plotkin constituents of $\RM(r,m)$
are readily visualized using a binary tree, called a \emph{Plotkin
tree}, and denoted $\Tree{r}{m}$.  Each vertex in the tree is labelled
with a pair of integers $(r',m')$; in particular, the root is labelled
with $(r,m)$.  A vertex $v$ with label $(r',m')$ where $m'>0$ has two
children: the left child has label $(r',m'-1)$
and the right child has label $(r'-1,m'-1)$.
Vertices $(r',m')$ with $m'=0$ are leaf vertices; they have no children.
Fig.~\ref{fig:tree} illustrates $\Tree{2}{4}$.

Since it is possible for some vertices in $\Tree{r}{m}$ to have the same
label, it is useful to assign a unique variable-length binary
\emph{address} to each vertex.  The construction is standard: the root
is assigned address $\varnothing$, and throughout the tree (or any
rooted subtree thereof), the left child of a vertex with the address
$\mathbf{a}$ receives address $(\mathbf{a} \mid 0)$, while the right
child receives address $(\mathbf{a} \mid 1)$.  However, rather than
writing addresses as vectors, we will write them as binary strings.  For
example, Fig.~\ref{fig::RM_address} illustrates the vertex addresses for
a rooted subtree of $\Tree{3}{6}$.  If the root of the tree has label
$(r,m)$, then a vertex with address $\mathbf{a}$ has label
$(r-\text{wt}(\mathbf{a}),m-\text{len}(\mathbf{a}))$, where
$\text{wt}(\mathbf{a})$ denotes the Hamming weight of $\mathbf{a}$ and
$\text{len}(\mathbf{a})$ denotes the length of $\mathbf{a}$.

In principle, soft-decision ML decoding of $\RM(r,m)$ can be
accomplished via the Viterbi algorithm; however the number of edges in
the trellis diagram scales at least exponentially in the block length
\cite[Sec.~IV.A]{Forney_CosetCodesII}, making this approach
computationally infeasible for all but very short RM codes.  On the
other hand, ML decoding of the trivial RM codes (which include
repetition codes and SPC codes) is easy.  ML decoding of $\RM(m-1,m)$, the 
SPC code, is straightforward via the \emph{Wagner}
decoding rule~\cite{SPC-decoder}, with the main complexity being
determination of the position in the received word having the smallest LLR
magnitude.   Soft-decision decoding of first-order RM codes $\RM(1,m)$
can be accomplished using the ``Green Machine''
decoder~\cite[Ch.~14]{Macwilliams-book}\cite{FHT-decoder}, with the main
complexity being the computation of a Hadamard transform of the received
LLR vector.  In the following sections, we discuss some of the non-ML
soft-decision decoding algorithms for general RM codes.

\subsection{GMC Decoding of RM Codes}

As already noted, the GMC decoder \cite{GMC} uses a divide-and-conquer
approach that exploits the recursive Plotkin structure of RM codes,
splitting the task of decoding $\RM(r,m)$ into that of decoding its
Plotkin constituents, namely $\RM(r-1,m-1)$ and $\RM(r,m-1)$.  In turn,
these smaller decoding problems can themselves be split into even
smaller decoding problems, until eventually an ``easy'' decoding problem
is reached, at which point no further task-splitting is required.

To make this precise, let $\mathcal{A}$, called an \emph{atom set}, be
any collection of RM codes for which task-splitting is not required due
to the availability of computationally feasible decoders for those
codes.  For example, $\mathcal{A}$ might contain all trivial RM codes
and the first-order RM codes
$\RM(1,m)$.  We will assume that an atom set always contains trivial
RM codes $\RM(r,0)$ of length 1.  An RM code \emph{not} contained in
$\mathcal{A}$ is called \emph{composite} with respect to $\mathcal{A}$.
The decoding function for an RM code $\RM(r,m) \in \mathcal{A}$ is
denoted as $A_{r,m}$.

Let $\bm{\lambda} \in \XR^{2^m}$, denote the LLR
vector produced at the output of a binary-input memoryless channel when
a codeword of $\RM(r,m)$ is transmitted.  The GMC decoder with atom set
$\mathcal{A}$ is the function $\GMC{r}{m}^{\mathcal{A}} :
\XR^{2^m} \to \RM(r,m)$ recursively defined by
\begin{equation}
\GMC{r}{m}^{\mathcal{A}}(\bm{\lambda}) \triangleq
\begin{cases}
A_{r,m}(\bm{\lambda}), & \text{if }\RM(r,m) \in \mathcal{A};\\
(\mathbf{u} \mid \mathbf{u} \oplus \mathbf{v}), & \text{otherwise},\end{cases} 
\label{eq::GMC concat}
\end{equation}
where, assuming $\bm{\lambda}$ is partitioned as
$\bm{\lambda}=(\bm{\lambda'} \mid \bm{\lambda''})$ with $\bm{\lambda'},
\bm{\lambda''} \in \XR^{2^{m-1}}$, the binary
$2^{m-1}$-vectors $\mathbf{v}$ and $\mathbf{u}$ are given as
\begin{align}
\mathbf{v} &= \GMC{r-1}{m-1}^{\mathcal{A}} (\bm{\lambda'} \boxplus \bm{\lambda''}),
\label{eq::GMC_LLRv}
\\
\mathbf{u} &= \GMC{r}{m-1}^{\mathcal{A}} (\bm{\lambda'} + (-1)^{\mathbf{v}} \bm{\lambda''}),
\label{eq::GMC_LLRu}
\end{align}
respectively.  Since $\mathcal{A}$ is required to include trivial
length-1 RM codes, the GMC decoder is guaranteed to terminate.

Note that computation of $\mathbf{u}$ in (\ref{eq::GMC_LLRu}) depends on
the availability of $\mathbf{v}$ defined in (\ref{eq::GMC_LLRv}), and
both $\mathbf{u}$ and $\mathbf{v}$ are required in (\ref{eq::GMC
concat}).  In effect, the GMC decoder $\GMC{r}{m}^{\mathcal{A}}$
performs a reverse post-order traversal of a rooted subtree of
$\Tree{r}{m}$---called the \emph{GMC decoding tree} with respect to
$\mathcal{A}$ and denoted as $\Tree{r}{m}^{\mathcal{A}}$---in which
vertices with labels corresponding to codes that are composite with
respect to $\mathcal{A}$ are internal, and those with labels
corresponding to elements of $\mathcal{A}$ are leaves.  We refer to the
decoders for elements of $\mathcal{A}$ as \emph{leaf decoders}.

\begin{figure}[t]
    \centering
    \includegraphics[scale=0.666667]{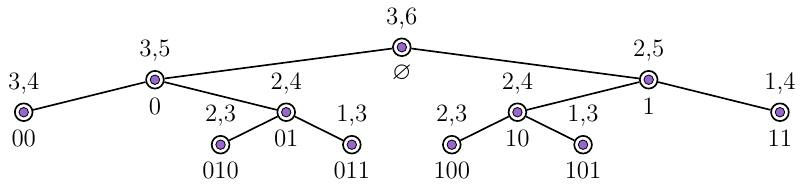}
    \caption{The address of each vertex in
$\Tree{3}{6}^{\mathcal{A}^{\ast}}$ is indicated below that vertex; the
vertex label is indicated above.}
    \label{fig::RM_address}
\end{figure}

The smallest possible atom set contains only the RM codes of length 1,
in which case the GMC decoding tree for $\RM(r,m)$ is the same as the
Plotkin tree $\Tree{r}{m}$.  However, as noted earlier, another choice
for the atom set is the one containing the trivial and the first-order
RM codes, \emph{i.e.},
\begin{equation}
\mathcal{A}^{\ast} = \left\{ \RM(r, m): m \in \mathbb{Z}^{\geq 0}, r \in
\mathbb{Z}^{\leq 1} \cup \mathbb{Z}^{\geq m-1} \right\}.
\label{eq:atomset}
\end{equation}
The circled vertices in Fig.~\ref{fig:tree} illustrate
$\Tree{2}{4}^{\mathcal{A}^{\ast}}$, the GMC decoding tree for $\RM(2,4)$
with atom set $\mathcal{A}^{\ast}$, and Fig.~\ref{fig::RM_address} shows
the GMC decoding tree $\Tree{3}{6}^{\mathcal{A}^{\ast}}$ (along with the
address of each vertex).

It is important to note that GMC decoding, though computationally
convenient, does not always return an ML codeword, and therefore GMC
decoding does not generally have the smallest possible BLER.  The complexity
of GMC decoding of an RM code of length $n$ is $\mathcal{O}(n \log
n)$~\cite{GMC, recur-RM-decoder}. Because of its low complexity (and its
suboptimality), the GMC decoder can be used as an elementary decoder in
AE decoding; see Sec.~\ref{subsec::ae}.

\subsection{SCL Decoding of RM Codes}

SCL decoders generalize GMC decoders by allowing the constituent
decoders to return not just a single codeword, but rather a \emph{list}
of possible codewords.
In SCL decoding, the atom set is usually chosen as
$\mathcal{A}' \triangleq \left\{\RM(r,0): r \in \mathbb{Z}\right\}$,
which contain only RM codes of length 1.
To avoid an exponential growth of the decoding
list, an upper bound on the returned list size at each constituent
decoder is maintained, retaining only those candidate codewords of
lowest \emph{cost}, defined as follows \cite{LLR-based-SCL}.
For any LLR vector $\bm{\lambda} \in \XR^{2^m}$ and any
$\mathbf{c} \in \RM(r,m)$, let
\begin{equation}
\cost(\mathbf{c},
\bm{\lambda}) \triangleq \sum_{i \in [2^m]}
\lsigmoid\left((-1)^{\mathbf{c}[i]} \bm{\lambda}[i]\right)
\end{equation}
be the cost of decoding the LLR vector $\bm{\lambda}$
to the vector $\mathbf{c}$, where $\lsigmoid$ was defined
in (\ref{eq:lsigmoid}).
If $\bm{\lambda} =
(\bm{\lambda'} \mid \bm{\lambda''})$, for some $\bm{\lambda'}$ and
$\bm{\lambda''} \in \XR^{2^{m-1}}$, and if $\mathbf{c} =
(\mathbf{u} \mid \mathbf{u} \oplus \mathbf{v})$, then
\begin{align}
\cost(\mathbf{c}, \bm{\lambda}) =
\cost(\mathbf{v}, \bm{\lambda'} \boxplus \bm{\lambda''})+
\cost(\mathbf{u}, \bm{\lambda'} + (-1)^{\mathbf{v}} \bm{\lambda''}).
\label{eq:costdecomp}
\end{align}
The terms
$\cost(\mathbf{v}, \bm{\lambda'} \boxplus \bm{\lambda''})$ and 
$\cost(\mathbf{u}, \bm{\lambda'} + (-1)^{\mathbf{v}} \bm{\lambda''})$ may
be regarded as the
\emph{Plotkin constituent costs} of
$\mathbf{c} = (\mathbf{u} \mid \mathbf{u} \oplus \mathbf{v})$;
the SCL decoder exploits this additive decomposition of the overall
cost of $\mathbf{c}$.

SCL decoding proceeds recursively in the same manner as the GMC
decoder, except, rather than being provided with a single LLR vector,
the constituent decoders are provided with a list of one or more ordered pairs,
where each pair comprises an LLR vector and an associated cost.
The output of each constituent decoder is a list of triples,
where each triple comprises
a codeword, the index within the input list
of the LLR vector from which that codeword
was decoded, and the overall cost of that codeword.

More precisely, a constituent SCL decoder for
$\RM(r,m)$ with
maximum output-list-size $\lmax \in \mathbb{Z}^{\geq 2}$
and input-list-size $\lin \leq \lmax$
is a function
\[
\SCL_{r,m}^{\lin, \lmax}: (\XR^{2^m} \times
\XR)^{\lin} \to (\RM(r,m)  \times
[\lin] \times \XR)^{\lout},
\] 
where $\lout = \min(2^{k(r,m)}\lin, \lmax)$.
We will start by defining this function for the
special case when $m=0$, \emph{i.e.}, for a leaf decoder.

When $m=0$, the SCL decoder is a map
\[
\SCL_{r,0}^{\lin, \lmax}: (\XR \times \XR)^{\lin}
 \to (\RM(r,0) \times [\lin] \times \XR)^{\lout},
\]
where $\lout = \min(2^{k(r,0)}\lin, \lmax)$.
Thus the input to a leaf decoder is
a list of pairs
$((\lambda_0,\psi_0), \ldots, (\lambda_{\lin -1},\psi_{\lin-1})) \in
(\XR \times \XR)^{\lin}$, where $\lambda_i$ is an
LLR value and $\psi_i$ is the associated cost.
Two cases must be considered:
\begin{itemize}
\item If $r \geq 0$, then $\RM(r,0) = \{ \mathbf{0}_1, \mathbf{1}_1 \}$.
In this case,
for each $\lambda_i$, $i \in [\lin]$, there are two
possible codewords: $\mathbf{0}$ with overall cost
$ \psi_i + \cost(\mathbf{0}, \lambda_i) = \psi_i + \lsigmoid(\lambda_i)$
leading to the tentative output triple $(\mathbf{0},i,\psi_i + \lsigmoid(\lambda_i))$,
and $\mathbf{1}$ with overall cost $ \psi_i + \cost(\mathbf{1}, \lambda_i) = \psi_i + \lsigmoid(-\lambda_i)$
leading to the tentative output triple $(\mathbf{1},i,\psi_i + \lsigmoid(-\lambda_i))$.
In total there will be $2\lin$ tentative output triples.  If needed,
this list of tentative triples
is truncated to length $\lmax$, with the decoder
returning only those
candidates with the least overall cost.
\item If $r < 0$, then $\RM(r,0) = \{ \mathbf{0}_1 \}$. In this case $\mathbf{0}$ is the only
possible candidate codeword for each
$\lambda_i$, $i\in [\lin]$,
leading to output triple $(\mathbf{0},i,\psi_i + \lsigmoid(\lambda_i))$.
All $\lin$ of the resulting triples are returned by the decoder.
\end{itemize}

When $m>0$, suppose the input to the SCL decoder
is the $\lin$-tuple 
\[
\bm{\tau} = \left(\left((\bm{\lambda'}_i \mid \bm{\lambda''}_i), s_i\right)_{i \in
[\lin]}\right),
\] 
with $\bm{\lambda'}_i$ and
$\bm{\lambda''}_i\in \XR^{2^{m-1}}$  
and $s_i \in \XR$. We will then have
\begin{equation}
\SCL_{r,m}^{\lin, \lmax}(\bm{\tau}) \triangleq
\left(\left((\mathbf{u}_i
\mid \mathbf{u}_i \oplus \mathbf{v}_{q_i}), p_{q_i}, s'_i\right)_{i \in [\lout]}\right)
\label{eq:SCLrecursion}
\end{equation}
where
$\mathbf{v}_j$'s and the
$p_j$'s in (\ref{eq:SCLrecursion}) are obtained from 
\begin{equation}
\left(\left(\mathbf{v}_j, p_j, s_j''\right)_{j\in [\ell']}\right) = 
\SCL_{r-1, m-1}^{\lin, \lmax}\left(
(\bm{\lambda'}_i \boxplus \bm{\lambda''}_i,s_i)_{i \in [\lin]}
\right),
\label{eq::general_v}
\end{equation}
where 
\begin{equation}
\ell' = \min(2^{k(r-1, m-1)}\lin, \lmax).
\label{eq::ellp}
\end{equation}  
Moreover, the
$\mathbf{u}_j$'s, the $q_j$'s, and the $s'_j$'s in (\ref{eq:SCLrecursion}) are
obtained from
\begin{equation}
\left(\left(\mathbf{u}_j, q_j, s'_j\right)_{j\in [\ell'']}\right)
= 
\SCL_{r, m-1}^{\ell', \lmax}\left(
\big(\bm{\lambda'}_{p_j} + (-1)^{\mathbf{v}_j} \bm{\lambda''}_{p_j}, s_j''\big)_{j \in [\ell']}
\right),
\label{eq::general_u}
\end{equation}
where 
\begin{equation}
\ell'' = \min(2^{k(r, m-1)}\ell', \lmax) = \min(2^{k(r,m)} \lin, \lmax).
\label{eq::ellpp}
\end{equation}

The overall SCL decoder for $\RM(r,m)$ is obtained by setting $\lin =
1$ and by choosing any arbitrary real number as the initial cost of
its input LLR vector.  The overall decoder returns the codeword
having the smallest analog weight with respect to the input LLR vector
from the list of candidates returned by
the decoder at the root of the decoding tree.

\subsection{AE Decoding of RM Codes}\label{subsec::ae}

\begin{figure}
    \centering
     \includegraphics[scale=0.66666]{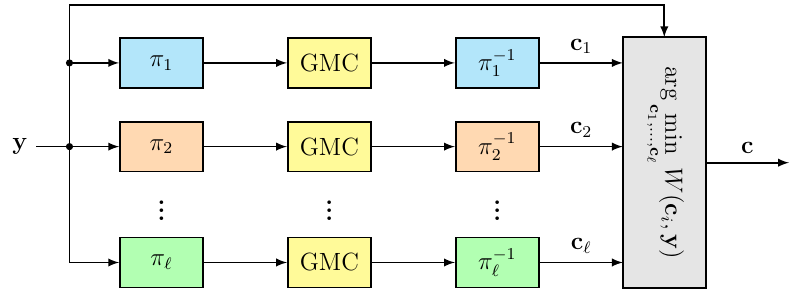}
    \caption{The general architecture of an AE decoder corresponding to
an automorphism ensemble of size $\ell$.}
    \label{fig::aut_dec}
\end{figure}

As described in Sec.~\ref{sec::intro} and as illustrated in
Fig.~\ref{fig::aut_dec}, an AE decoder operates by permuting the received
LLR vector $\bm{\lambda}$ by the elements of a set $\mathcal{E} = \{
\pi_1, \ldots, \pi_{\ell} \}$, called an \emph{automorphism ensemble}, of
code automorphisms.   Each of the $\ell$ permuted LLR vectors is then
decoded (perhaps in parallel) using a simple sub-optimal \emph{elementary
decoder} such as a GMC decoder, and the corresponding inverse permutation
is applied to each decoding result.  The most likely codeword is then
selected from the $\ell$ generated candidates.

For any choice of atom set $\mathcal{A}$ and
any automorphism $\pi \in \Aut{r, m}$, let
$\permGMC{r}{m}{\pi, \mathcal{A}} : \XR^{2^m} \to \RM(r,m)$ denote the
\emph{$\pi$-permuted GMC decoder}, defined as
\begin{equation}
\permGMC{r}{m}{\pi, \mathcal{A}} \triangleq 
\pi^{-1} \circ \GMC{r}{m}^{\mathcal{A}} \circ\, \pi,
\end{equation} 
where $\circ$ denotes function composition.
Thus $\permGMC{r}{m}{\pi,\mathcal{A}}(\bm{\lambda}) =
\pi^{-1}\GMC{r}{m}^{\mathcal{A}}(\pi \bm{\lambda})$.
Of course $\permGMC{r}{m}{\id, \mathcal{A}} = \GMC{r}{m}^{\mathcal{A}}$,
\emph{i.e.}, the $\id$-permuted GMC decoder is just the GMC decoder itself.

Depending on the choice of $\pi$, because
of the suboptimality of a GMC decoder, it may happen
that
\begin{equation}
\GMC{r}{m}^{\mathcal{A}} \neq \permGMC{r}{m}{\pi, \mathcal{A}}.
\label{eq:x4855u}
\end{equation}
This is useful, because if
$\GMC{r}{m}^{\mathcal{A}}$ makes an error, there is a possibility
that
$\permGMC{r}{m}{\pi, \mathcal{A}}$ can nevertheless decode correctly, and
\emph{vice versa}.
Automorphism ensemble decoding of RM codes~\cite{kamenev2019anew,
AE-SC,ivanov2019permutation} exploits (\ref{eq:x4855u}) to decode a received
LLR vector using an ensemble $\mathcal{E}$ of automorphisms with the property,
that
\begin{equation}
\pi_1, \pi_2 \in \mathcal{E},~
\pi_1 \neq \pi_2 \text{ implies }
    \permGMC{r}{m}{\pi_1, \mathcal{A}} \neq \permGMC{r}{m}{\pi_2, \mathcal{A}}.
 \label{eq::XT5612}
\end{equation}
Indeed, if $ \permGMC{r}{m}{\pi_i, \mathcal{A}}$, $\pi_i \in \mathcal{E}$, produces 
decoding decision $ \mathbf{c}_i $, then
the overall AE decoding decision is given, as illustrated
in Fig.~\ref{fig::aut_dec}, by
\[
\mathbf{c} =
\arg \min_{\mathbf{c}_1,\ldots,\mathbf{c}_{\ell}} w(\mathbf{c}_i,\bm{\lambda}),
\]
where the analog-weight function $w$ was
defined in (\ref{eq::analog_weight}).

\section{Constituent Automorphism Decoding} \label{sec::CA}

Like all successive-cancellation decoding schemes, GMC decoders are prone
to error propagation: once a leaf decoder makes an error, the error
propagates to all succeeding decoders encountered in the decoding
tree traversal.  Because of channel
polarization~\cite{channel-polarization}, different constituent
leaf decoders of an RM code observe different synthetic channels
and, as a result, they
have different \emph{first-error probability}, the probability that they
produce an error while all preceding leaf decoders have decoded
correctly.
For example, Table~\ref{tab::first_err} gives the binary
addresses of the leaf
constituent codes of $\GMC{4}{9}^{ \mathcal{A}^{\ast} }$ having the largest
first-error probability at the output
of a BI-AWGN channel with an SNR of $4$~dB.  It is clear that most decoding
errors are triggered by the leaf decoder with address 111.

Because of this channel polarization effect,
it is sensible to devote
more decoding resources to those constituent codes having relatively
degraded channels, as these trigger the most decoding errors.
However, a conventional AE decoder treats all constituent
codes evenly.  In particular, via an AE decoder with some automorphism ensemble
$\mathcal{E}$, each constituent code of a composite RM code is decoded
$|\mathcal{E}|$ times. To remedy this issue, we introduce
constituent automorphism
decoding.
The key idea is
to apply AE decoding \emph{locally} to the codes that are composite
with respect to a given atom set $\mathcal{A}$.

\begin{table}
\centering
\caption{Leaves with largest first-error probability in
$\GMC{4}{9}^{\mathcal{A}^{\ast}}$.}
\begin{tabular}{c|c|c|c}
address & contribution & address & contribution\\
\hline
$111$ & $91.6 \%$  & $1101$ & $7.39 \%$ \\
$11001$ & $0.5 \%$ & $1011$ & $0.4 \%$ \\\hline
\end{tabular}\\[1ex]
(BI-AWGN channel, SNR=4dB)
\label{tab::first_err}
\end{table}

With a random selection of $\mathcal{E}$ from $\Aut{r, m}$ chosen to satisfy
(\ref{eq::XT5612}), we did not see any difference in BLER at the output of a
BI-AWGN between various selected automorphism ensembles for any of the
$\RM(r,m)$ codes that we tested.  Thus we speculate that the performance of an
AE decoder depends only on the ensemble size.  As a result, we will only
specify the AE size, without describing the actual automorphisms used.

By default, all decoders in the decoding tree will use an AE size of one
(conveniently implemented with an automorphism ensemble containing just the
identity permutation $\id$).  However, certain selected composite (non-leaf)
decoders will use AE decoding locally with a larger AE size.

To specify the local decoders having an AE size greater than unity, we define
an \emph{automorphism distribution} $\mathcal{S}$ as a set of $(\mathbf{a},s)$
pairs, where $\mathbf{a}$ is the binary address of a decoder in
$\Tree{r}{m}^{\mathcal{A}}$ and $s$ is the corresponding AE size.  For example,
$\mathcal{S} = \{ (1,2), (11,3) \}$ denotes a CA decoder that applies AE
decoding with AE size 2 at the decoding tree vertex with address 1, and AE
decoding with AE size 3 at the vertex with address 11.  In this notation, a
conventional AE decoder with AE size $s$ has automorphism distribution
$\mathcal{S} = \{ (\varnothing,s) \}$. We denote the CA decoder for $\RM(r,m)$
with automorphism distribution $\mathcal{S}$ and atom set $\mathcal{A}$ as
\[\CA_{r, m}^{\mathcal{S}, \mathcal{A}}: \XR^{2^m} \to \RM(r, m).\]

In the rest of this paper, all CA decoders use the atom set $\mathcal{A}^\ast$.
The leaf decoders associated with $\mathcal{A}^\ast$ are permutation invariant ML
decoders, so applying AE to them cannot improve performance. Hence the AE size
of leaf nodes in $\Tree{r}{m}^{\mathcal{A}^\ast}$ is fixed to unity. As we will
show in Sec.~\ref{sec::results}, most of the benefit of CA decoding arises when
the addresses $\mathbf{a}$ are chosen from the set $\{ \varnothing, 1, 11, 111,
\ldots \}$ which are called \emph{rightmost nodes}, as these correspond to the
synthetic channels that polarize to relatively ``bad'' channels.  A slightly
better heuristic for choosing the vertices at which to apply AE decoding is
given in Sec.~\ref{subsec::selection}.

\section{Complexity Analysis}
\label{sec::complexity}

\subsection{Basic Operations}

It is difficult to give a precise analysis of decoding complexity, as it
strongly depends on the available hardware and its architecture.  Nevertheless,
complexity (measured by the area-time product of an integrated circuit
implementation, or in execution time of a software implementation) will scale
approximately linearly with the number of unary and binary operations executed
by the decoding algorithm.  Accordingly, we will measure
complexity by counting such operations (in the worst case) for each of the
decoders that we consider.

In practice, LLR values are often represented using
a fixed point (integer) representation with a fixed word size.
We will assume that
vectors of length $n$, whether integer- or $\mathbb{F}_2$-valued,
are stored as sequences of $n$ words.
All decoders that we consider are implemented as sequences of the following
basic unary and binary operations (operating on
one or two words): addition ($+$), comparison ($<$, $>$, $\leq$, $\geq$), 
minimum and maximum ($\min$, $\max$),
soft XOR ($\boxplus$), negative log-sigmoid ($\lsigmoid$), absolute value ($|\cdot|$), 
negation ($-$), binary addition ($\oplus$), and, finally, copying a word or its negation.

We assume the soft XOR and the negative log-sigmoid operations are calculated by
look-up tables or hardware-friendly approximations.
In a pipelined processor architecture, the fetching of operands
and the storing of results take place in parallel with
instruction decoding and execution, so we do not
assign additional complexity for such operations.
Assuming the use of pipelining,
all of these basic operations can be executed in a single clock cycle on 
average.  Accordingly, we weight these operations evenly, taking
their total count $\rchi$ as a measure of decoder complexity.

Let $D_{r,m}$ denote a decoder for $\RM(r,m)$ and let $\rchi_{D}(r,m)$ denote
its complexity.  Except at leaf decoders, $D_{r,m}$ is defined recursively in
terms of $D_{r-1,m-1}$ and $D_{r,m-1}$, the decoders for its Plotkin
constituents.   Depending on the nature of the decoder, these constituent
decoders may be invoked more than once.  It follows that $\rchi_D(r,m)$ will
generally decompose (recursively) into three terms: \romannumeral 1) a term
that accounts for one or more applications of $D_{r-1,m-1}$, \romannumeral 2) a
term that accounts for one or more applications of $D_{r,m-1}$, and
\romannumeral 3) a term that accounts for the preparation of LLRs and the final
decision.  Leaf decoders are \emph{not} defined recursively, and thus each such
decoder will require its own separate complexity analysis.

\subsection{Complexity of CA, GMC, and AE decoders}

For all nontrivial RM codes,
assuming use of the atom set $\mathcal{A}^{\ast}$ defined in (\ref{eq:atomset}),
the only required leaf decoders are those for the
$\RM(m-1,m)$ SPC codes and $\RM(1,m)$ first-order
RM codes.

\subsubsection{$\RM(m-1,m)$}
The worst-case complexity of applying the Wagner decoding rule
to the $\RM(m-1,m)$ SPC code can be separated into:
\begin{itemize}
\item converting LLRs to hard-decisions, which
requires $2^m$ comparisons;
\item determining the overall parity, which
requires $2^m-1$ binary field additions;
\item checking whether the overall parity is nonzero,
which requires $1$ comparison,
\item flipping the bit
with the least absolute LLR
value (if the parity bit is nonzero), which
requires $2^m$ absolute value computations,
$2^m-1$ comparisons, and $1$ binary field addition.
\end{itemize}
Adding these values gives
\begin{equation}
    \rchi(m-1,m) = 2^{m+2}. \label{eq::complexity of SPC}
\end{equation}

\subsubsection{$\RM(1,m)$}
The worst-case complexity of decoding $\RM(1,m)$, the
length-$2^m$ first-order RM code,
using the Green Machine decoder can be separated into:
\begin{itemize}
\item applying the fast Hadamard transform to
the given LLR vector, which requires $m \cdot 2^m$ additions;
\item finding the index with the largest absolute value, which
requires $2^m$ absolute value computations, and $2^m-1$ comparisons;
\item extracting the sign of the corresponding LLR,
which requires $1$ comparison,
\item converting the sign and index to a codeword,
 which requires
$2^m + m$ operations according to Algorithm~\ref{algo:RMcodeword}.
\end{itemize}
Adding these values gives
\begin{equation}
    \rchi(1,m) = (m+3)2^m +m \label{eq::complexity of 1st RM} .
\end{equation}

\begin{algorithm}[t]
\caption{Map sign and index to $\mathbf{v} \in \RM(1,m)$.}
\label{algo:RMcodeword}
\begin{algorithmic}[1]
\Require $m \in \mathbb{Z}^{\geq 0}$, sign $s \in \mathbb{F}_2$, index vector $\mathbf{i} \in \mathbb{F}_2^m$
\State $\mathbf{v} \gets  (s)$ 
\For {$j\in [m]$}
\If{$\mathbf{i}[m-1-j] = 0$}
  \State $\mathbf{v} \gets (\mathbf{v} \mid \mathbf{v})$
  \Comment ($2^j$ copy operations.)
\Else
  \State $\mathbf{v} \gets (\mathbf{v} \mid \overline{\mathbf{v}})$
  \Comment ($2^j$ negated-copy operations.)
\EndIf
\EndFor\\
\Return $\mathbf{v}$
\end{algorithmic}
\emph{Remarks:}
For each $j \in [m]$,
executing line 4 or 6 requires $2^j$ copy or negated-copy operations.
Neglecting loop overhead,
we have one initial copy of the sign
bit, $1$ comparison per loop (for a total of $m$ comparisons)
and $\sum_{i=0}^{m-1} 2^i = 2^m-1$ copy operations, yielding
a total complexity of $2^m + m$ operations.
\end{algorithm}

Let $\rchi_{\CA}(r, m, \mathcal{S})$ denote the complexity of $\CA_{r,
m}^{\mathcal{S}, \mathcal{A}^\ast}$, the CA decoder for a composite $\RM(r,m)$
under an automorphism distribution $\mathcal{S}$ and with atom set
$\mathcal{A}^\ast$. Let $\ell$ denote the number of automorphisms applied at
the root of the decoding tree. Note that $(\varnothing, \ell) \in \mathcal{S}$
if and only if $\ell > 1$.  A CA decoder for $\RM(r,m)$ using $\mathcal{S}$
invokes CA decoders for the Plotkin constituents of the code,
\emph{i.e.}, $\RM(r-1, m-1)$ and $\RM(r, m-1)$, using, respectively, the
automorphism distributions $\mathcal{S}_1$ and $\mathcal{S}_0$, defined as
\begin{align*}
\mathcal{S}_1 &\triangleq \left\{(\mathbf{a}, s_{\mathbf{a}}): \big((1 \mid
\mathbf{a}), s_{\mathbf{a}}\big) \in \mathcal{S}\right\} \text{ and}\\
\mathcal{S}_0 &\triangleq \left\{(\mathbf{a}, s_{\mathbf{a}}): \big((0 \mid
\mathbf{a}), s_{\mathbf{a}}\big) \in \mathcal{S}\right\}.
\end{align*}

The number of operations contributing to $\rchi_{\CA}(r, m, \mathcal{S})$ decomposes as follows:
\begin{itemize}
\item preparing LLRs for $\CA_{r-1,m-1}^{\mathcal{S}_1, \mathcal{A}^\ast}$ in accordance with (\ref{eq::GMC_LLRv}), which
requires $\ell 2^{m-1}$ soft XOR operations;
\item invoking $\ell$ instances of $\CA_{r-1, m-1}^{\mathcal{S}_1, \mathcal{A}^\ast}$, which requires 
$\ell \rchi_{\CA}(r-1, m-1, \mathcal{S}_1)$ operations;
\item preparing LLRs for $\CA_{r, m-1}^{\mathcal{S}_0, \mathcal{A}^\ast}$ in accordance with (\ref{eq::GMC_LLRu}), which
requires $\ell 2^{m-1}$ comparisons and $\ell 2^{m-1}$ additions;
\item invoking $\ell$ instances of
$\CA_{r, m-1}^{\mathcal{S}_0, \mathcal{A}^\ast}$, which requires 
$\ell \rchi_{\CA}(r, m-1, \mathcal{S}_0)$ operations;
\item combining Plotkin constituents according
to (\ref{eq::GMC concat}), which requires
$\ell 2^{m-1}$ binary field additions; 
\item choosing the codeword with minimum analog weight among $\ell$ candidates, which if $\ell = 1$ 
requires no operations and  
if $\ell > 1$ requires $\ell 2^m$ comparisons
and $\ell (2^{m} - 1)$ additions for computing analog weight, and 
another $\ell - 1$ comparisons for choosing the final codeword.
\end{itemize}
As a result, 
\begin{align*}
\rchi_{\CA}(r, m, \mathcal{S}) =  &\ell \rchi_{\CA}(r-1, m-1, \mathcal{S}_1) + \ell \rchi_{\CA}(r, m-1, \mathcal{S}_0) + \\ 
&\ell 2^{m+1} + \mathbbm{1}_{\ell > 1}\cdot (\ell 2^{m+1} - 1),
\end{align*}
with boundary conditions $\rchi_{\CA}(m-1, m, \{\}) = \rchi(m-1, m)$ and
$\rchi_{\CA}(1,m, \{\}) = \rchi(1,m)$, defined in (\ref{eq::complexity of SPC})
and (\ref{eq::complexity of 1st RM}), respectively.

GMC and AE decoders are special cases of a CA decoder and so
their complexity may be expressed in terms of 
$\rchi_{\CA}$.
Since all nodes in the decoding tree of a GMC decoder use only
the identity automorphism,
the automorphism distribution for such a decoder
is $\mathcal{S} = \{\}$; thus
\begin{equation}
\COMP{GMC}(r,m) = \rchi_{\CA}(r, m, \{\}).
\end{equation}
An AE decoder with an automorphism ensemble of size $\ell$ has automorphism 
distribution $\mathcal{S}=\{(\varnothing, \ell)\}$; thus,
\begin{equation}
\COMP{AE}(r, m, \ell) = \rchi_{\CA}(r, m, \{(\varnothing, \ell)\}).
\end{equation}

\subsection{Complexity of SCL Decoders}

We assume that $\RM(r,m)$ has a non-negative order and that the atom set
$\mathcal{A}'$ is used.  Similar to CA decoders,
we first discuss the complexity of list decoding of codes in
$\mathcal{A}'$. We then derive the complexity of the SCL
decoder for a composite code $\RM(r,m)$.

\subsubsection{$\RM(r,m) \in \mathcal{A}'$}

For list-truncation purposes, some of the leaf SCL decoders must find the least
$\lmax$ numbers among an unsorted list of $n$ numbers, where $n \in
\mathbb{Z}$.
When $\lmax < n$, there are a number of algorithms for this task,
each having its own complexity.
We denote by
$\COMP{sel}(\lmax,n)$ the number of basic
operations required for this task
in the worst case.  Clearly $\COMP{sel}(\lmax,n) = 0$
when $\lmax \geq n$.

In this paper, we use $(\lmax,n)$
\emph{selection
networks}, which combine a fixed number of \emph{comparator} and
\emph{minimum selector} (minselector) elements to achieve a deterministic number of basic
operations for
selection of the smallest $\lmax$ entries from
an input list of size $n > \lmax$~\cite[Sec.~5.3.4]{knuth1998theart},
\cite{yao1975astudy}. 
A comparator is a two-input two-output computational unit, that
transforms
a pair $(x,y)$ of input scores to $(\min(x,y),\max(x,y))$, requiring
two basic operations.
Similarly, a minselector is a two-input one-output unit,
transforming a pair $(x,y)$ to $\min(x,y)$, requiring one basic operation.
Since $n-\lmax$
inputs are dropped by an $(\lmax,n)$ selection network, the number of
minselectors in such a network will always be $n-\lmax$.
Thus, an $(\lmax,n)$ selection network with a total
of $T$
computational units (comparators
and minselectors) requires 
$\COMP{sel}(\lmax,n) = 2T - (n-\lmax)$ basic operations.

\begin{figure}
\centering
\begin{tabular}{cc}
\includegraphics[scale=0.625]{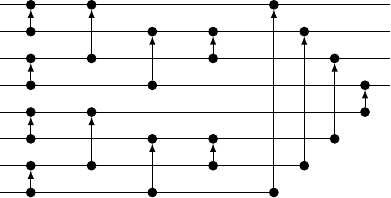} & 
\includegraphics[scale=0.625]{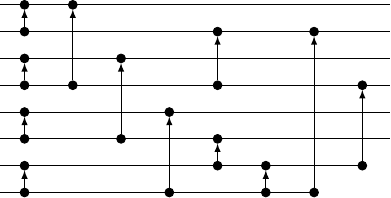}\\
(a) & (b)
\end{tabular} 
\caption{Selection networks with
parameters: (a)~(4,8), (b)~(6,8).
Vertical segments denote comparators (having two output ports)
and minselectors (having one output port), with
an arrow pointing towards the output port corresponding to
the minimum input.}
\label{fig::selection network}
\end{figure}

An $(\lmax,n)$ selection network is said to be \emph{optimal}
if the total number $T$
of computational units
that it contains is as small as possible;
this
minimum possible number of computational units is denoted as
$U_{\lmax}(n)$.
The exact value of $U_{\lmax}(n)$ is
known only for
small values of $\lmax$ and $n$; however, it is known
that~\cite[Sec.~5.3.4, Thm.~A]{knuth1998theart}
\begin{equation}
U_{\lmax}(n) \geq (n-\lmax) \lceil \log_2 (\lmax+1) \rceil.
\label{eq::alekseev}
\end{equation}
Fig.~\ref{fig::selection network} shows conjectured-optimal $(4,8)$
and optimal $(6,8)$ selection networks.

Table~\ref{tab::heap} shows the value for $\COMP{sel}(\lmax,n)$
for different
values of $\lmax$ and $n$ encountered in SCL decoding of RM codes
in Sec.~\ref{sec::results}.  The values for
$\COMP{sel}(4,8)$ and $\COMP{sel}(6,8)$ correspond to
the selection networks of
Fig.~\ref{fig::selection network}.
In the $(6,12)$ case, 
we have assumed that the lower bound in (\ref{eq::alekseev})
is achieved with equality;  however, in practice this bound may not
be achievable, and thus $\COMP{sel}(6,12)$ may be larger
than the value given in the table.

\begin{table}[b]
\centering
\caption{Values for $\COMP{sel}(\lmax,n)$}
\begin{tabular}{c|c|c|c|c|c}
$\lmax,n$ & $\COMP{sel}$ & 
$\lmax,n$ & $\COMP{sel}$ & 
$\lmax,n$ & $\COMP{sel}$\\
\hline
4,8 & 24 & 
6,8 & 22 &
6,12 & 30\\
\hline
\end{tabular}
\label{tab::heap}
\end{table}

Let $\rchi_{\SCL}(r, m, \lin, \lmax)$ denote the decoding complexity of
$\SCL_{r,m}^{\lin, \lmax}$.  We start by considering
$\rchi_{\SCL}(r, 0, \lin, \lmax)$, \emph{i.e.}, the decoding complexity
of the leaf decoders $\SCL_{r,0}^{\lin, \lmax}$. 

For an input-list size $\lin \in \mathbb{Z}^{\geq 1}$ and a maximum output-list
size $\lmax \in \mathbb{Z}^{\geq 2}$, the complexity of applying
$\SCL_{r, 0}^{\lin, \lmax}$ to decode LLR values $\lambda_0, \ldots,
\lambda_{\lin - 1}$ to a codeword (of length one)
in $\RM(r,0)$ can be separated into
the following steps.
\begin{itemize}
\item Computing the overall cost of each tentative candidate codeword
along with the value of that codeword.
For  the zero codeword this requires 
one call to $\lsigmoid$, one addition, and  one bit-copy operation,
for a total cost of $3\lin$ basic operations.
When $r \geq 0$,
the nonzero codeword requires these operations and
an extra negation for an additional cost of $4 \lin$ basic operations.
In total,
this results in $\lin (3+4 \cdot \mathbbm{1}_{r \geq 0})$ operations.
\item Finding the (at most) $\lmax$
codewords having least cost from
a list of $2^{k(r,0)}\lin$ codewords, which
requires $\COMP{sel}(\lmax,2^{k(r,0)}\lin)$ basic operations.
\end{itemize}
Adding the number of basic operations for these steps gives
\begin{align}
\rchi_{\SCL}(r, 0, \lin, \lmax) =
\lin (3+4 \cdot \mathbbm{1}_{r \geq 0}) + \COMP{sel}(\lmax,2^{k(r,0)}\lin).
\label{eq::scl_boundary}
\end{align}

\subsubsection{$\RM(r,m) \not\in \mathcal{A}'$}
For an input-list size $\lin$ and a maximum
output-list size $\lmax$, the complexity of applying $\SCL_{r,m}^{\lin, \lmax}$
to decode $\RM(r,m) \not\in \mathcal{A}'$ decomposes as follows:
\begin{itemize}
\item preparing LLRs for $\SCL_{r-1, m-1}^{\lin, \lmax}$ in accordance with (\ref{eq::general_v}), which
requires $\lin 2^{m-1}$ soft XOR operations; 
\item invoking $\SCL_{r-1, m-1}^{\lin, \lmax}$, which
requires $\rchi_{\SCL}(r-1, m-1, \lin, \lmax)$ operations;
\item preparing LLRs in accordance with (\ref{eq::general_u}), which
requires $\ell' 2^{m-1}$ comparisons 
and $\ell' 2^{m-1}$ additions where $\ell'$ is given in (\ref{eq::ellp});
\item invoking $\SCL_{r, m-1}^{\ell', \lmax}$, which  requires $\rchi_{\SCL}(r, m-1, \ell', \lmax)$ operations;
\item combining Plotkin constituents, which
requires $\ell'' 2^{m-1}$ binary field additions where $\ell''$   
is as (\ref{eq::ellpp}).
\end{itemize}
As a result,
\begin{multline}
\rchi_{\SCL}(r, m, \lin, \lmax) = \rchi_{\SCL}(r-1, m-1, \lin, \lmax) + \\
\rchi_{\SCL}(r, m-1, \ell', \lmax) + 2^{m-1}(\lin + 2\ell' + \ell''), 
\end{multline}
with boundary conditions defined by leaf decoders according to (\ref{eq::scl_boundary}).
The complexity of the overall SCL decoder with a maximum output-list size of
$\lmax$ is then $\rchi_{\SCL}(r, m, 1, \lmax)$.

\section{Numerical Results} \label{sec::results}

In this section, using numerical simulations we provide
performance versus
complexity and BLER versus SNR trade-offs for the
schemes described in this paper.
We assume transmission over a BI-AWGN channel.  Performance is measured by
the gap to the constrained Shannon limit at a BLER of $10^{-3}$.  Complexity
is measured as the worst-case number of operations,
computed according to Sec.~\ref{sec::complexity}, normalized by
the code dimension.

\subsection{Selection of automorphism distribution}
\label{subsec::selection}

There are numerous possible ways to choose an
automorphism distribution.  It is an open problem
to determine the automorphism distribution that leads
to the best performance at a given decoding complexity.
Guided by the following numerical
results, we provide a heuristic method for finding 
``good'' CA decoders.

\begin{figure}
    \centering
    \includegraphics[scale=0.6667]{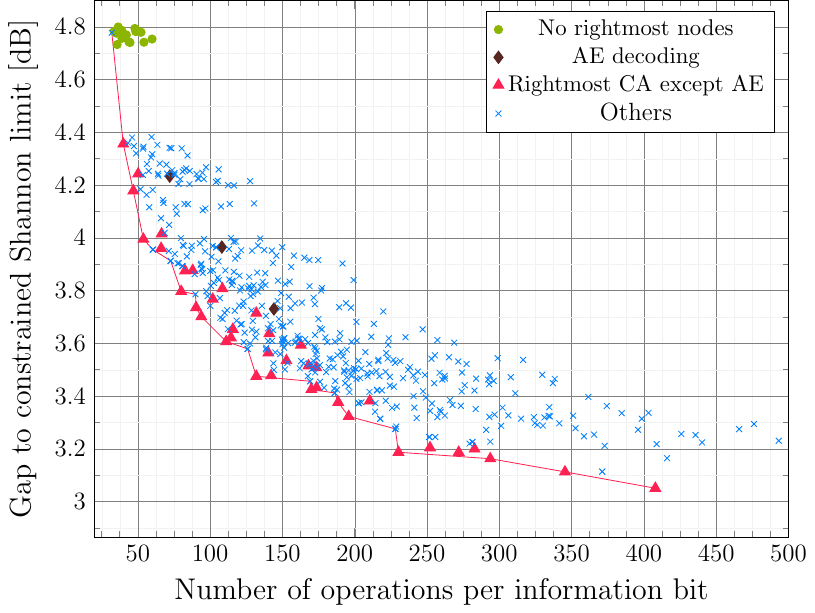}
    \caption{An extensive test of CA decoders of RM$(4,9)$ at
BLER=$10^{-3}$. Candidates nodes are those with address in
the set $\{ \varnothing, 1, 11, 110, 1100 \}$ with ensemble size for
each candidate node ranging from $1$ to $4$.}
    \label{fig::exhaustive}
\end{figure}

We first compare the performance-complexity tradeoffs of CA decoders
for $\RM(4,9)$ for a large number of different choices of
automorphism distribution.  
According to
Table~\ref{tab::first_err}, most decoding errors are caused by
rightmost leaf decoders.
We have tested addresses in automorphism distributions
involving the first $5$ nodes in the reverse pre-order decoding
tree traversal (namely, nodes with address
$\varnothing$, 1, 11, 110, and 1100),  with AE size for each address
ranging from $1$ to $4$.
The results are given in Fig.~\ref{fig::exhaustive}.

The lower hull
in Fig.~\ref{fig::exhaustive} is the \emph{Pareto frontier}; it connects
the \emph{Pareto-efficient points}. A point is Pareto-efficient
with respect to a given set of points
if no other point in the set
can achieve better performance with the same or lower complexity.
The parameters of the Pareto-efficient points in this test are
given in Table~\ref{tab::exhaust_eff}. 
\begin{table}[]
\centering
\caption{Pareto-efficient points in Fig.~\ref{fig::exhaustive}.}
\begin{tabular}{c|c|c}
Decoder & Operations per information bit & Gap to CSL \\ \hline
GMC & 32.043 & 4.778\\ \hline
$(11,2)$ & 39.984 & 4.356\\ \hline
$(11,3)$ & 46.93 & 4.178\\ \hline
$(11,4)$ & 53.875 & 3.995\\ \hline
$(11,4);(1100,2)$ & 60.469 & 3.955\\ \hline
$(1,2);(11,2);(1100,2)$ & 72.734 & 3.913\\ \hline
$(1,2);(11,3)$ & 80.031 & 3.797\\ \hline
$(1,2);(11,3);(1100,2)$ & 89.922 & 3.786\\ \hline
$(1,3);(11,2)$ & 90.301 & 3.736\\ \hline
$(1,2);(11,4)$ & 93.922 & 3.701\\ \hline
$(1,3);(11,3)$ & 111.137 & 3.607\\ \hline
$(1,3);(11,3);(1100,2)$ & 125.973 & 3.579\\ \hline
$(1,3);(11,4)$ & 131.973 & 3.475\\ \hline
$(1,4);(11,2);(110,3)$ & 169.18 & 3.456\\ \hline
$(1,4);(11,4)$ & 170.023 & 3.425\\ \hline
$(1,4);(11,3);(110,2)$ & 186.258 & 3.412\\ \hline
$(\varnothing,2);(1,3);(11,2)$ & 188.598 & 3.376\\ \hline
$(\varnothing,2);(1,2);(11,4)$ & 195.84 & 3.323\\ \hline
$(\varnothing,2);(1,4);(110,3)$ & 228.105 & 3.276\\ \hline
$(\varnothing,2);(1,3);(11,3)$ & 230.27 & 3.187\\ \hline
$(\varnothing,3);(1,2);(11,4)$ & 293.762 & 3.162\\ \hline
$(\varnothing,3);(1,3);(11,3)$ & 345.406 & 3.113\\ \hline
$(\varnothing,3);(1,3);(11,4)$ & 407.914 & 3.051\\ \hline
\end{tabular}

\label{tab::exhaust_eff}
\end{table}

We categorize the tested automorphism distributions  in
Fig.~\ref{fig::exhaustive} into four classes: (a) no rightmost nodes, (b) only
root node, (c) only rightmost nodes (excluding only root node), (d) others.
It is clear that CA decoders in class (a) have almost the same decoding
performance as the GMC decoder, which implies that including the rightmost nodes
is a necessary condition to make an improvement. The Pareto frontier has roughly
$0.2$~dB gain from decoders in (b), which shows the advantage of CA decoding
compared to AE decoding. CA decoders with automorphism distributions satisfying
(c) are called \emph{rightmost} CA decoders. According to
Fig.~\ref{fig::exhaustive} and Table~\ref{tab::exhaust_eff}, the
Pareto-efficient points involve automorphism distributions where only
rightmost or ``nearly rightmost'' vertices in the decoding
tree receives an AE size greater than one.

These observations suggest the following heuristic method
for choosing an automorphism distribution where only
rightmost composite decoders (\emph{i.e.}, the root or any
non-leaf decoder with an all-ones address) receive an AE size greater
than unity.  Let $m(i)$ denote the AE size for the right-most
composite decoder at depth $i$ in $\Tree{r}{m}^{\mathcal{A}^*}$.
The heuristic would then choose:
\begin{enumerate}
    \item $m(i) \leq 7$  for all $i$, \emph{i.e.},  the number of
automorphisms applied at any local decoder is restricted; and
    \item if $m(i) \geq 2$ then $m(j) \geq 2$ for all $j\geq i$, \emph{i.e.},
AE decoding with size greater than unity should be
applied at a node only
when all of its right descendants also have AE size greater
than unity.
\end{enumerate}

\begin{figure}
    \centering
    \includegraphics[scale=0.66667]{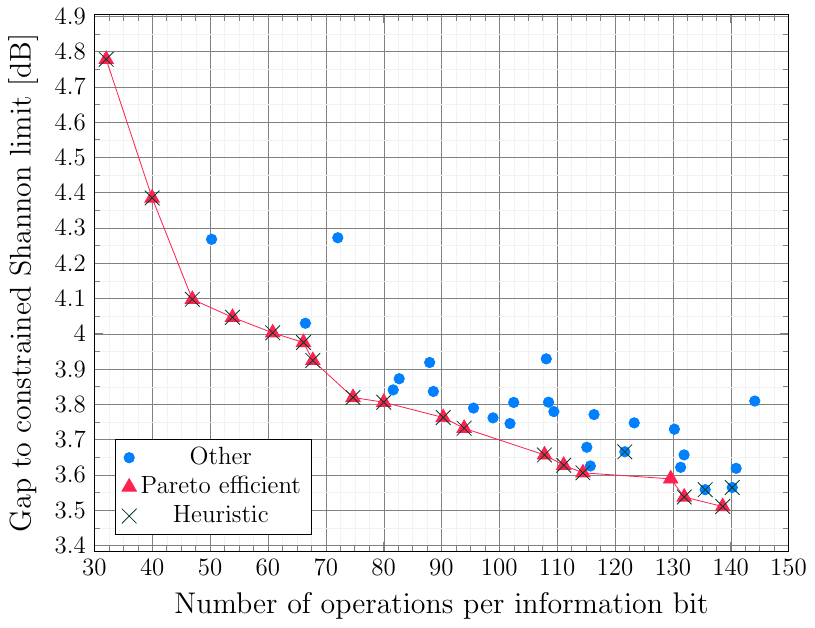}
    \caption{CA decoders for code RM$(4,9)$ at BLER=$10^{-3}$ having
complexity no more than AE-$4$.}
    \label{fig::all rightmost}
\end{figure}

Fig.~\ref{fig::all rightmost} shows the performance of all decoders
for $\RM(4,9)$, where AE decoding is applied at the right-most
composite decoders.  Included are those decoders 
whose complexity does not exceed that of AE-4, \emph{i.e.},
the CA decoder with $\mathcal{S} = \{(\varnothing,4) \}$.
It can be seen that
almost all Pareto-efficient points are selected by
the given heuristic and all heuristically chosen decoders are close to
the Pareto frontier (within $0.07$~dB). The decoder parameters
for Pareto-efficient points in Fig.~\ref{fig::all rightmost}
are given in Table~\ref{tab::heuristic}.

In the following comparisons, 
the CA decoders were all selected using the heuristic described above.

\begin{table}[]
    \centering
    \caption{Pareto-efficient points in Fig.~\ref{fig::all rightmost}.}
    \label{tab::heuristic}
    \begin{tabular}{c|c|c}
Decoder & Operations per information bit & Gap to CSL \\ \hline
GMC & 32.043 & 4.778\\ \hline
$(11,2)$ & 39.984 & 4.385\\ \hline
$(11,3)$ & 46.93 & 4.097\\ \hline
$(11,4)$ & 53.875 & 4.046\\ \hline
$(11,5)$ & 60.82 & 4.003\\ \hline
$(1,2);(11,2)$ & 66.141 & 3.975\\ \hline
$(11,6)$ & 67.766 & 3.924\\ \hline
$(11,7)$ & 74.711 & 3.819\\ \hline
$(1,2);(11,3)$ & 80.031 & 3.806\\ \hline
$(1,3);(11,2)$ & 90.301 & 3.762\\ \hline
$(1,2);(11,4)$ & 93.922 & 3.732\\ \hline
$(1,2);(11,5)$ & 107.812 & 3.656\\ \hline
$(1,3);(11,3)$ & 111.137 & 3.627\\ \hline
$(1,4);(11,2)$ & 114.461 & 3.606\\ \hline
$(\varnothing,2);(11,5)$ & 129.637 & 3.589\\ \hline
$(1,3);(11,4)$ & 131.973 & 3.537\\ \hline
$(1,5);(11,2)$ & 138.621 & 3.51\\ \hline
\end{tabular}

\end{table}

\subsection{Comparison of Decoders}

\begin{figure}
    \centering
\setlength{\tabcolsep}{0em}
\begin{tabular}{c}
\includegraphics[scale=0.6666]{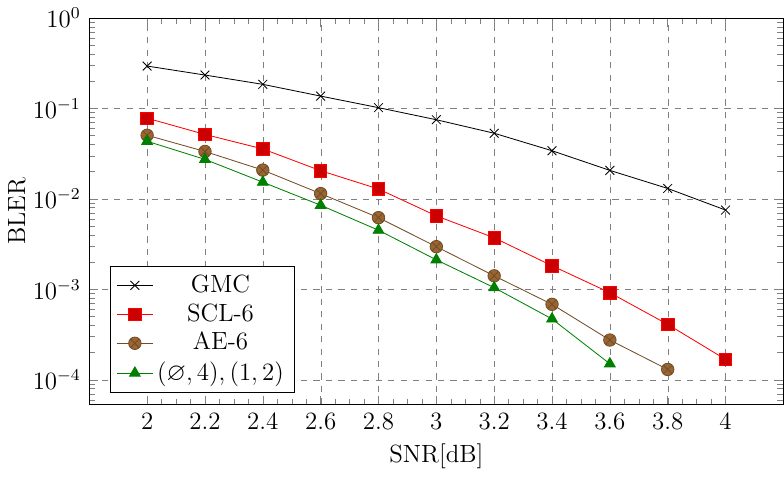}\\
(a)\\
\includegraphics[scale=0.6666]{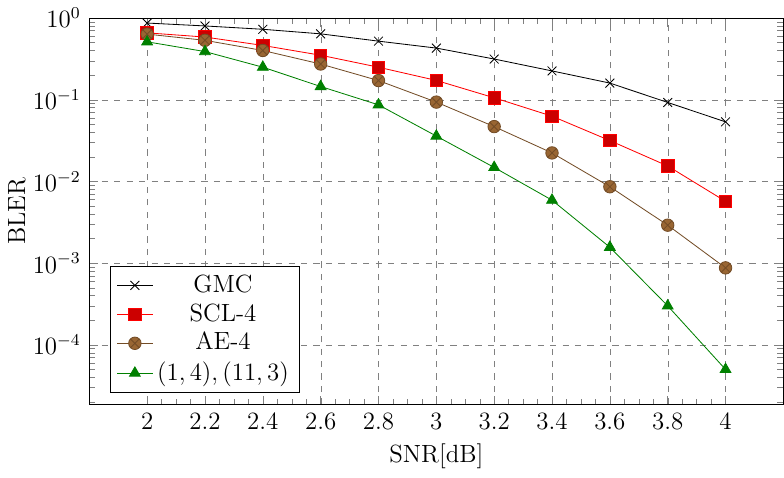}\\
(b)\\
\includegraphics[scale=0.6666]{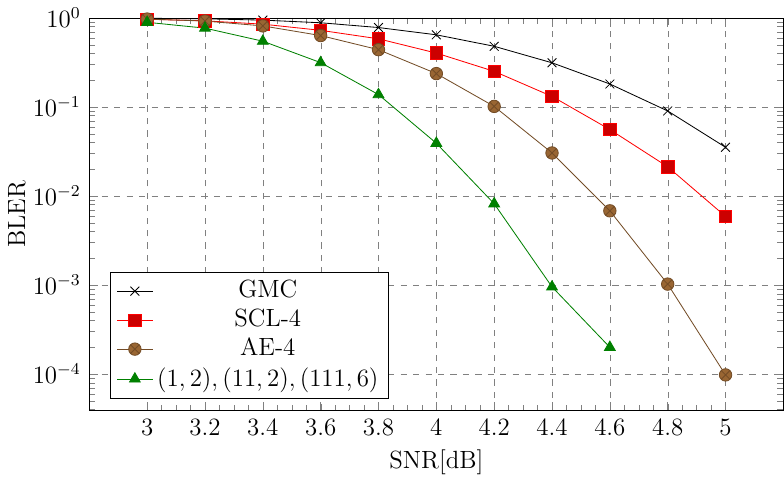}\\
(c)
\end{tabular}
\caption{Performance of decoders for RM codes of rate 1/2:
(a)~$\RM(3,7)$, (b)~$\RM(4,9)$, and (c)~$\RM(5,11)$.}
\label{fig::half rate performance}
\end{figure}

Fig.~\ref{fig::half rate performance} compares the performance
achieved by different decoders for three different RM
codes of rate 1/2. Table~\ref{tab::half rate comp} lists the complexity
for each tested decoder. The AE, CA and SCL decoders that we have tested are
approximately at the same complexity level. It is clear that the CA
decoding algorithm outperforms the other decoders at the same complexity
level.

\begin{table}
\centering
\caption{Decoding complexity for RM codes of rate 1/2 in Fig.~\ref{fig::half rate performance}.}
\begin{tabular}{c|c|c}
Code & Decoder & Operations per information bit \\ 
\hline
\multirow[c]{4}{*}[0in]{$\RM(3,7)$} & GMC & 25.10\\ 
\cline{2-3} 
& SCL-$6$ & 225.80 \\ 
\cline{2-3}
& AE-$6$    & 174.55\\ 
\cline{2-3} 
& $(\varnothing,4),(1,2)$ & 175.05\\ 
\hline
\multirow[c]{4}{*}[0in]{$\RM(4,9)$} & GMC & 32.04\\ 
\cline{2-3}
& SCL-4 & 195.47\\ 
\cline{2-3}
& AE-4    & 144.17\\ 
\cline{2-3} 
& $(1,4),(11,3)$ & 142.24 \\ 
\hline
\multirow[c]{4}{*}[0in]{$\RM(5,11)$} & GMC & 39.15 \\ 
\cline{2-3}
& SCL-4 & 230.31\\ 
\cline{2-3}
& AE-4 & 172.6\\ 
\cline{2-3} 
& $(1,2),(11,2),(111,6)$ & 157.41 \\ 
\hline
\end{tabular}

\label{tab::half rate comp}
\end{table}

Comparing the performance of CA and AE decoding, we find that the
improvement changes for different codes. The improvement for
$\RM(3,7)$ is merely $0.1$~dB, whereas there is a gain of
$0.4$~dB for $\RM(5,11)$.

\begin{figure}
    \centering
\setlength{\tabcolsep}{0em}
\begin{tabular}{c}
\includegraphics[scale=0.6666]{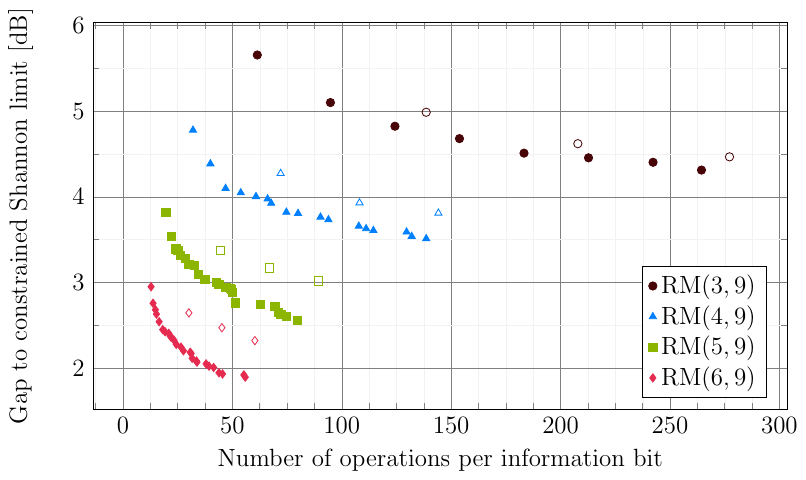}\\
(a)\\
\includegraphics[scale=0.66667]{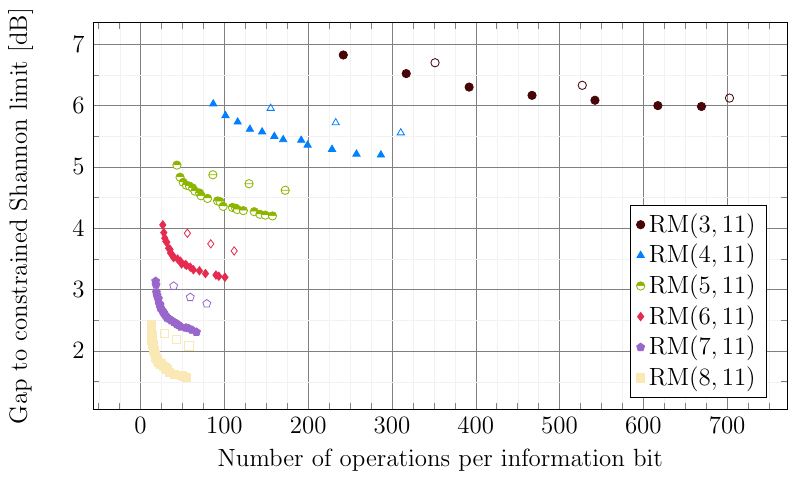}\\
(b)\\
\end{tabular}
\caption{Performance of RM codes with the same $m$ (a) $m=9$,
(b) $m=11$.}
\label{fig::same length}
\end{figure}

Fig.~\ref{fig::same length} shows a comparison of CA decoders and AE
decoders for different RM codes with the same block length (same $m$).
The Pareto-efficient CA decoders with decoding
complexity not exceeding that of AE-$4$ are depicted with solid marks,
while the AE decoders are drawn in
hollow marks. We observe that the gap between the AE decoder and the
CA decoder is larger for higher order $r$.
In other words, the gain achieved when applying CA
decoding compared to AE decoding is more significant for high order RM
codes. 
This trend is expected, since
higher order RM codes have more rightmost nodes, which means there
are more nodes observing ``bad'' channel polarization, and therefore
such codes benefit more from an uneven distribution of decoding
resources.

\section{Conclusions}\label{sec::conclusions}

This paper has introduced the constituent automorphism (CA) decoding
algorithm for RM codes, which applies AE decoding at the (mainly rightmost)
constituent decoders according to a specified automorphism distribution.
CA decoders can achieve better performance versus
complexity trade-offs than other state-of-the-art decoding algorithms.
The benefits of CA decoding appear to increase as the order of
the RM code increases.
While the problem of selecting an automorphism distribution
achieving the best performance at a fixed complexity remains open,
we have provided a simple heuristic by which a ``good'' automorphism
distribution may be selected.
In the future, it would be interesting to study whether CA
decoding can be adapted to a suitable class of polar codes. 
As in \cite{polar-aut-design}, 
the chosen polar-code class would need to overcome
the property that polar codes do not generally have a large
automorphism group \cite{not-many-aut-polar}.

\IEEEtriggeratref{7}
\bibliographystyle{IEEEtran}
\bibliography{IEEEabrv,ref}

\end{document}